\documentclass[amsmath,amssymb,nofootinbib]{revtex4}
\usepackage{dcolumn}
\usepackage{lipsum}
\usepackage{bm}
\usepackage{graphicx}
\usepackage{color}
\usepackage{amsmath}
\newcommand{\be}{\begin{equation}}
\newcommand{\ee}{\end{equation}}
\newcommand{\ba}{\begin{eqnarray}}
\newcommand{\ea}{\end{eqnarray}}

\newcommand{\n}[1]{\label{#1}}
\newcommand{\eq}[1]{Eq.(\ref{#1})}

\newcommand{\hhh}{\, ,\hspace{0.2cm}}

\newcommand{\hu}{\widehat{U}}
\newcommand{\hw}{\widehat{W}}
\newcommand{\hx}{\widehat{X}}
\newcommand{\hv}{\widehat{V}}

\begin{document}

\title{A Distorted Charged Dilaton Black Ring} 

\author{Shohreh Abdolrahimi${}^{a,b}$ and Christos C. Tzounis${}^{a,c}$}
\affiliation{${}^a$Department of Physics and Astronomy, California State Polytechnic University, 91768 CA, USA\\
${}^b$Kavli Institute for Theoretical Physics, UC Santa Barbara, 93106 CA, USA\\
${}^c$Schmid College of Science and Technology, Chapman University, 92866 CA, USA }

\date{2022 June 9}

\begin{abstract}
We present the distorted five dimensional static black rings in Einstein-Maxwell-dilaton (EMd) gravity by using the sigma model structure of the dimensionally reduced field equations. We investigate how a static and neutral distribution of external matter distorts the black ring. In the case of a distorted five dimensional static black rings in EMd gravity the distortion parameters can be adjusted such that the solution is free of conical singularities in the region of the validity of the solution.  
\end{abstract}

\maketitle

\section{Introduction}
This paper is an extension of our previous results \cite{Abdolrahimi2020} to a charged case. 
Black holes and black objects interact with external matter and fields. We can gain a a full understanding of the general theory of relativity and the properties of black holes only if we study the interaction of black holes with matter and sources. Many exact solutions of Einstein equations represent idealized situations, i.e. a black hole in an empty universe. There are several ways to study interacting black holes. The best way of for studying interaction of the black holes is the numerical analysis and the study of the dynamical systems. However, considerable insight can be gained from studying exact (or approximate) solutions describing a black hole tidally distorted by external matter. In particular, our goal is to see which properties of black holes/black objects are more general and remain intact even for interacting black holes. 

Geroch and Hartle \cite{GerochH} proposed to consider static (or stationary), axisymmetric space-times which are not asymptotically flat that model to some extend the interaction of a black hole with the external matter. We consider a ``local" black hole. Such solution is valid only in the neighborhood of a black hole and is distorted by external sources, located far away. The external matter is not included into the solution and ``located" at the asymptotic infinity. In what follows, we use the word distorted black hole in this context.  Analysis of the distorted black holes have the major advantage that these solutions are valid for broad classes of external matter, although we may need to impose regularity condition as well as some symmetry restrictions.  

In the case of static and vacuum, axisymmetric space-times, the external metric near  distorted black holes is constructed using the Weyl form \cite{Weyl}. Four-dimensional, distorted, static, axisymmetric, vacuum black holes were studied in, e.g., \cite{dis1,dis2,dis3,dis4,dis5}. Distorted, static, axisymmetric, electrically charged black holes were studied in, e.g., \cite{Fairhurst,Israel1776,Abdolrahimi2009}, and distorted, stationary, axisymmetric, vacuum and electrically charged black holes were studied in, e.g., \cite{Tomimatsu,Breton1997} and \cite{Breton1998,Ansorg2008,Ansorg2009,Ansorg2009-2}, respectively. 

The construction of the distorted black hole solutions to higher dimensions one relies on generalized Weyl form of the metric \cite{GeneralizedWeyl}. The first higher dimensional distorted black hole was constructed in \cite{Abdolrahimi2010}. Distorted five-dimensional electrically charged black holes and distorted Myers-Perry black holes have been respectively constructed in \cite{Abdolrahimi2013, Abdolrahimi2014}. Distorted and deformed black holes have been constructed and studied in various other papers \cite{SCN3, dis2A, dis4A, Poisson:2009qj, Semerak:2016gfz,  Shoom:2015rda, Basovnik:2016awa, Ansorg:2010ru, Shoom:2015slu, Abdolrahimi:2014msa,Abdolrahimi:2017pmt, Kunz:2017mfe}.

The study of distorted black holes have manifested some strange and remarkable properties. For sufficiently distorted Schwarzschild solutions, the isolated horizon can be foliated with neither marginally trapped nor outer trapping two-dimensional surfaces \cite{Pilkington}. The distortion can cause the space-time curvature to become very high at some regions of black hole horizon \cite{dis5,Abdolrahimi2009}. In the case of the distorted Myers-Perry black hole, the ratio of the horizon angular momentum and the mass $J^2/M^3$ is unbounded, and can grow arbitrarily large \cite{Abdolrahimi2014}. Similarly, in contrast to the case of the undistorted Kerr black hole, for a distorted Kerr black hole $J^2/M^4>1$ for some distortion parameters \cite{Abdolrahimi:2015c}. In the case of octupole distortions of a Schwarzschild black hole the local observer may see two images or eyebrow-like structure from a single black hole \cite{Abdolrahimi:2015a,Abdolrahimi:2015b,Grover:2018tbq}. These features illustrate the importance of study of distorted black holes in the investigation of the properties of black holes and illustrating which properties of black holes are unique and universal. In higher dimensions, there exist other black hole solutions that do not have spherical horizons, such as black rings, black saturns, di-rings, and bicycling black rings, black helical rings, as well as more general blackfolds \cite{ER3,Emparan2008,ER2, Elvang:2007rd,Iguchi, Elvang2008, Izumi2008, Evslin:2007fv,Emparan:2011br}. In our perspective, it is necessary to investigate the features of distorted versions of these black objects to fully be able to know which properties of these black holes or black objects are universal. Various studies were devoted to investigating how the properties of 4-dimensional isolated black holes are influenced if they are distorted by an external matter field, and which of them remain unaffected, such as the analysis of the algebraic type of a distorted black hole or the local first law of thermodynamics for distorted black holes \cite{Papadopoulos1984, Ashtekar1999,Ashtekar2000,Ashtekar2000v2}. 

In a recent paper, we constructed a five-dimensional static black ring distorted by a static and neutral distribution of external matter \cite{Abdolrahimi2020}. We demonstrated that the gravitational field of these external sources can be adjusted to remove the conical singularity of the undistorted black ring solution. The properties of the distorted black ring for the specific cases of dipole and quadrupole distortions was analyzed. In this paper we construct a charged black ring solution of Einstein-Maxwell-dilaton (EMd) gravity in 5-dimensions, distorted by a static and neutral distribution of external matter. This paper is organized as follows: In Sec. II, we discuss the generating technique for the construction of the solution. In Sec. III, we first overview the black ring solution, then apply the generating technique to construct the distorted charged black ring solution. In Sec. IV we briefly analyze some of the properties of the solution and summarize our results. In this paper, we use the following convention of units: $G(5) = c = 1$, the spacetime signature is +3, and the sign conventions are those adopted in \cite{MTW} .

\section{Generating Technique}
The Einstein-Maxwell-dilaton (EMd) gravity in $D$-dimensional spacetimes is described by the action \cite{Sen,Gibbons,Garfinkle}

\begin{equation}
S= {1\over 16\pi} \int d^Dx \sqrt{-g}\left(R - 2g^{\mu\nu}\partial_{\mu}\varphi \partial_{\nu}\varphi  -
e^{-2\alpha\varphi}F^{\mu\nu}F_{\mu\nu} \right), 
\end{equation}
where $R$ is the Ricci scalar, $\alpha$ is the dilaton coupling constant, $F_{\mu\nu}=\nabla_\mu A_\nu- \nabla_\nu A_\mu$, is the electromagnetic field tensor, and $\varphi$ is the dilaton field. The field equations derived from the action are
\begin{eqnarray}
R_{\mu\nu} &=& 2\partial_{\mu}\varphi \partial_{\nu}\varphi + 2e^{-2\alpha\varphi} \left[F_{\mu\rho}F_{\nu}^{\rho} - {g_{\mu\nu}\over 2(D-2)} F_{\beta\rho} F^{\beta\rho}\right], \\
\nabla_{\mu}\nabla^{\mu}\varphi &=& -{\alpha\over 2} e^{-2\alpha\varphi} F_{\nu\rho}F^{\nu\rho}, \\
&\nabla_{\mu}&\left[e^{-2\alpha\varphi} F^{\mu\nu} \right]  = 0 .
\end{eqnarray}
Here, $R_{\mu\nu}$ is the Ricci tensor. We consider static spacetimes and denote the Killing vector by $\xi$ ($\xi={\partial}/{ \partial t}$). In this section, we will restrict ourselves to generating asymptotically flat  solutions. Although, a distorted black hole solution is not asymptotically flat. The point is that in the absence of the distortion fields the solution reduces to the asymptotically flat EMd static black ring. We follow the approach presented in \cite{Yazadjiev}. Let us consider a static asymptotically flat  solution of $D$-dimensional Einstein equations
\begin{equation}
ds_{0}^2 = -e^{2U_{0}}dt^2 + e^{-{2U_{0}\over D-3}}h_{ij}dx^idx^j.  
\end{equation}
We consider the matrix corresponding to the target space following the procedure presented in \cite{Yazadjiev}
\begin{equation} P_{0}= e^{(\alpha_{D} -1)U_{0} }
\left(%
\begin{array}{cc}
  e^{2U_{0}} & 0 \\
0 & -1 \\\end{array}%
\right) , 
\end{equation}
and the metric $h_{ij}$. 
The action is invariant under the symmetry transformation which can be used to generate a solution of the $D$-dimensional EMd gravity given by the matrix
\begin{equation}
P = AP_{0}A^{T}.
\end{equation}
and with the same metric $h_{ij}$, where
\begin{equation}
A = \left(%
\begin{array}{cc}
  \cosh(\gamma) & \sinh(\gamma) \\
  \sinh(\gamma) &\cosh(\gamma) \\\end{array}%
\right). 
\end{equation}
Explicitly, we have, (for details see \cite{Yazadjiev}) 
\begin{eqnarray}
e^{U} &=& {e^{U_{0}}\over \left[\cosh^2(\gamma) - e^{2U_{0}}\sinh^2(\gamma) \right]^{1\over 1 + \alpha^2_{D} } }, \n{Trans1a}\\
e^{-\varphi_{D}} &=&  \left[\cosh^2(\gamma) - e^{2U_{0}}\sinh^2(\gamma) \right]^{\alpha_{D}\over 1 + \alpha^2_{D} }, \n{Trans2a}\\
\Phi_{D} &=& {\tanh(\gamma) \over \sqrt{1 + \alpha^2_{D}} } {1 - e^{2U_{0}}\over 1 - e^{2U_{0}}\tanh^2(\gamma) }. \n{Trans3a}
\end{eqnarray}
The static asymptotically flat black ring metric can be presented in a simple form:  
\be
ds_{0}^2=-\frac{F(x)}{F(y)}dt^2+\frac{1}{A^2(x-y)^2}\biggl[\biggl(\frac{F(y)^2}{1-x^2} dx^2+\frac{F(x) F(y)}{y^2-1}dy^2\biggl)+F(x)(y^2-1)d\psi^2+\frac{F(y)^2}{F(x)}(1-x^2) d\phi^2\biggl],\n{METRIC1}
\ee
where $F(x)=1-\mu x$ and $F(y)=1-\mu y$. The coordinate $y$ is in the range $y\leq -1$ with the black hole horizon located at $y\rightarrow -\infty$. The coordinate $x$ is in
the range $-1\leq x \leq 1$. The parameters $\mu$ and $A$ will be taken to lie in the range $0\leq\mu\leq1$, $A>0$. The EMd black ring solution was generated using SO(1, 1) transformations \cite{Yazadjiev}. The black ring is not free of conical singularity. This conical singularity can be placed either at $x=1$ or at $x=-1$. If the asymptotic metric does not contain a conical singularity, then the mass of the black ring is  
\be
M =\frac{3\pi\mu(1+\mu)}{4G_5 A^2}\,,
\ee
where $G_5$ is  Newton's constant in five dimensions. For  fixed  $A$, a change in $\mu$ changes  the black ring's mass. We can choose, the deficit membrane extended to infinity. In this case, the mass of the ring is 
\be
M=\frac{3\pi\mu\sqrt{1-\mu^2}}{4G_5 A^2}\,.
\ee
In what follows, the sign $\pm$ corresponds to taking the conical singularity at $x=\pm 1$, respectively. The area of the horizon is
\begin{equation} \n{areaeq1}
{\cal A}_{H\pm} = 8\pi^2  \cosh^{3\over 1 + \alpha^2_{5}}(\gamma) {\mu^2 \sqrt{(1+ \mu)(1 \pm \mu)}\over A^3}
\end{equation}
The transformations (\ref{Trans1a}-\ref{Trans3a}) were used to generate the following EMd solution \cite{Yazadjiev}, 
\begin{eqnarray}
ds^2 = - \left[\cosh^2(\gamma) - \sinh^2(\gamma){F(x)\over F(y)} \right]^{-2\over 1 + \alpha^2_{5}}{F(x)\over F(y)}dt^2 \\ + {\left[\cosh^2(\gamma) - \sinh^2(\gamma){F(x)\over F(y)} \right]^{1\over 1 + \alpha^2_{5}}\over A^2(x-y)^2} \left[ F(x)(y^2-1)d\psi^2  \right. \nonumber \\ \left. +  {F(x)F(y)\over y^2 -1}dy^2
 + {F^2(y)\over 1-x^2 }dx^2 + F^2(y){1-x^2\over F(x) }d\phi^2\right] ,\nonumber \\
e^{-\varphi_{5}} = \left[\cosh^2(\gamma) - \sinh^2(\gamma){F(x)\over F(y)} \right]^{\alpha_{5}\over 1 + \alpha^2_{5}} , \\
\Phi_{5} = {\tanh(\gamma)\over \sqrt{1 + \alpha^2_{5}} } {1 - {F(x)\over F(y) }\over 1 - \tanh^2(\gamma) {F(x)\over F(y)} } .
\end{eqnarray}

Initially, this solution was presented first in \cite{KL} without derivation. This solution was derived by the above generating transformation in \cite{Yazadjiev}. Detailed analysis of the solution is given in \cite{KL,Yazadjiev}. The solution is asymptotically flat. The charge and the temperature are given by
\begin{eqnarray}
&&Q_{\pm} = \pi \sqrt{{3\over 1 + \alpha^2_{5}}} \sinh(\gamma)\cosh(\gamma) { \mu\sqrt{(1+\mu)(1\pm \mu)} \over A^2 }, \\
&&T = {A\over 4\pi\mu \cosh^{3\over 1 + \alpha^2_{5}}(\gamma)}.
\end{eqnarray}
\section{Distorted Charged Black Ring}
\subsection{Seed Solution, a Distorted Black Ring}
Here, we will consider the distorted black hole ring solution recently generated in \cite{Abdolrahimi2020} given by
\ba
ds^2&=&-e^{2(\hu+\hw)}\frac{F(x)}{F(y)}dt^2+\frac{1}{A^2(x-y)^2}\biggl[e^{2(\hat{V}+\hu+\hw)}\biggl(\frac{F(y)^2}{1-x^2} dx^2+\frac{F(x) F(y)}{y^2-1}dy^2\biggl)\nonumber\\
&+&e^{-2\hw}F(x)(y^2-1)d\psi^2+e^{-2\hu}\frac{F(y)^2}{F(x)} (1-x^2)d\phi^2\biggl]~.\n{MetricA}
\ea
The distortion fields $\hu$ and $\hw$ satisfy the three-dimensional Laplace equation
\be\n{LaplaceEq}
\hx_{,\rho\rho}+\frac{1}{\rho}\hx{,\rho}+\hx_{,zz}=0, 
\ee
where $\hx$ is either $\hu$, or $\hw$. Let us consider the cylindrical coordinates $(\rho,z)$, related to $x$ and $y$ by the transformation
\ba
&&\rho=\frac{\alpha}{A^2(x-y)^2}\sqrt{F(x)F(y)(1-x^2)(y^2-1)}~,\n{Trans1}\\
&&z=\frac{\alpha(1-xy)(F(x)+F(y))}{2A^2(x-y)^2}~.\n{Trans2}
\ea
In this case, the solution of Eq. (\ref{LaplaceEq}) is well known and has the following form:
\ba\n{3.1}
\hx(\rho,z)=\sum_{n\geq0}\left[A_n\,R^{n}+B_n\,R^{-(n+1)}\right]P_n(\cos\vartheta)\,,
\ea
where \ba\n{3.2}
R=\frac{\sqrt{\rho^2+z^2}}{m}\hhh\cos\vartheta=z/R\,. 
\ea
Here, $P_n(\cos\vartheta)$ are the Legendre polynomials of the first kind. 
The coefficients $A_n$ and $B_n$ in the expansion \eq{3.1} correspond to interior and exterior multipole moments, respectively \cite{Multipoles1,Multipoles2}. We consider only the case of a black ring distorted by the presence of the external sources. In other words, we assume all $B_n$'s are zero. In what follows, for $\hu$, $A_n$'s are named $a_n$ and for $\hw$, $A_n$'s are named $b_n$.  We shall simply call the $a_n$ and $b_n$ coefficients {\em multipole moments}. One can use the analogy with the Newtonian picture in order to give an interpretation for multiple moments. Therefore, we expect that that higher-order multiple moments are expected to be smaller than the lower-order multiple moments. In section VI, we restrict our analysis to $n=0...2$. The case with $a_1\neq 0$ and $a_2=0$ is called the dipole distortion. The case with with $a_1= 0$ and $a_2\neq 0$ is called the quadrupole distortion. The case with $a_1\neq 0$ and $a_2\neq 0$ is called the dipole-quadrupole distortion.
If the distortion fields $\hu$ and $\hw$ are known, the function $\hv$ can be derived explicitly for  multiple moments $n=1..4$. For details see \cite{Abdolrahimi2020} 
\ba
&&\hat{V}=\hat{V}_1+\hat{V}_2,\n{formV1}\\
&&\hat{V}_1=\sum_{n,k\geq1}\frac{nk}{n+k}(a_na_k+a_nb_k+b_nb_k)R^{n+k}
[P_nP_k-P_{n-1}P_{k-1}]\,,\n{3.4c} 
\ea
\ba
&&\hv_2=\frac{1}{2z}\biggl[a_1{(R_1-2R_2+2R_3)} -b_1{(R_1+R_2-R_3)}-3(a_1+b_1)z \biggl]R P_1,~~~~~n=1,\\
\hv_2&=&-\frac{3}{2}(a_2+b_2)R^2P_2+\frac{1}{2m}\biggl[\biggl(a_2(R_1-2R_2+2R_3)-b_2(R_1+R_2-R_3)\biggl)\nonumber\\
&+&\frac{1}{z}\biggl(a_2{(R_1c_1-2R_2c_2+2R_3c_3)}-b_2{(R_1c_1+R_2c_2-R_3c_3)}\biggl)\biggl]R P_1,~~~~~n=2\\
\hv_2&=&-\frac{3}{2}(a_3+b_3)R^3P_3+\frac{1}{4z^3}\biggl(-a_3(R_1^3-2R_2^3+2R_3^3)+b_3(R_1^3+R_2^3-R_3^3)\biggl)R^3P_1^3\nonumber\\
&+&\frac{3}{4zm^2}\biggl[a_3\biggl ( (R_1-2R_2+2R_3)z^2+R_1c_1^2-2R_2 c_2^2+2R_3 c_3^2\biggl)\nonumber\\
&+&b_3\biggl(-(R_1+R_2-R_3)z^2-R_1c_1^2-R_2 c_2^2+R_3 c_3^2\biggl)\biggl]RP_1,~~~~~~n=3
\ea 
\ba
\hv_2&=& -\frac{3}{2} (a_4+b_4) R^4 P_4 -\frac{1}{z^3}\biggl( a_4 (R_1^3-2R_2^3+2R_3^3)-b_4(R_1^3+R_2^3-R_3^3)\biggl)R^4P_1^4 \nonumber\\
&+&\frac{1}{m^3}\biggl[\left(3a_4 (\frac{1}{2} R_1-R_2+R_3)-\frac{3b_4}{2}(R_1+R_2-R_3)\right)z^2\nonumber\\
&+&3\left(-a_4(\frac{1}{2}R_1 c_1-R_2 c_2+R_3 c_3) +\frac{b_4}{2}(R_1c_1+R_2 c_2-R_3 c_3)\right)z\nonumber\\
&+& \frac{a_4 r^2}{4 z}\left((R_1-2R_2+2R_3)z-R_1 c_1+2R_2 c_2-2R_3 c_3\right)\nonumber\\
&-& \frac{b_4 r^2}{4 z}\left((R_1+R_2-R_3)z-R_1 c_1-R_2 c_2+R_3 c_3\right)\nonumber\\
&+&\frac{1}{2 z} \left(a_4(R_1 c_1^3-2R_2 c_2^3+2 R_3 c_3^3)-b_4(R_1 c_1^3+R_2 c_2^3-R_3 c_3^3)\right)\nonumber\\
&+&\frac{3}{2}\left(a_4(R_1c_1^2-2R_2c_2^2+2R_3c_3^2)-b_4(R_1c_1^2+R_2 c_2^2-R_3 c_3^2)\right)\biggl] R P_1\,~~~~~~n=4\,. 
\ea
Here, $P_n\equiv P_n(z/R)$, and
\ba
&&\xi_i\equiv z-c_i~,\\
&&R_1\equiv \sqrt{\rho^2+\xi_1^2}~,\\
&&R_2\equiv \sqrt{\rho^2+\xi_2^2}~,\\
&&R_3\equiv -\sqrt{\rho^2+\xi_3^2}~,\\
&&Y_{ij}\equiv R_i R_j +\xi_i \xi_j +\rho^2~\\
&& c_1=1/(2A),~~~c_2=\mu/(2A),~~~c_3=-\mu/(2A). 
\ea

\subsection{Einstein-Maxwell-dilaton distorted black ring}
For the distorted black ring, we have 
\be
\exp(U_0)=\exp(\hu+\hw) \sqrt{\frac{F(x)}{F(y)}}\, 
\ee
Using the transformations Eqs. (\ref{Trans1a}-\ref{Trans3a}) we generate the metric of a static charged distorted black ring 
\ba
ds^2&=&-e^{2(\hu+\hw)}\frac{F(x)}{F(y)}  \left[\cosh^2(\gamma) - \exp[2(\hu+\hw)]\sinh^2(\gamma){F(x)\over F(y)} \right]^{-\frac{2}{1 + \alpha^2_{5}}}  dt^2 \nonumber\\
&+&\frac{1}{A^2(x-y)^2} {\left[\cosh^2(\gamma) - \exp[2(\hu+\hw)]\sinh^2(\gamma){F(x)\over F(y)} \right]^{1\over 1 + \alpha^2_{5}}}\biggl[e^{2(\hat{V}+\hu+\hw)}\biggl(\frac{F(y)^2}{1-x^2} dx^2+\frac{F(x) F(y)}{y^2-1}dy^2\biggl)\nonumber\\
&+&e^{-2\hw}F(x)(y^2-1)d\psi^2+e^{-2\hu}\frac{F(y)^2}{F(x)} (1-x^2)d\phi^2\biggl]~.\n{metricEMDL}
\ea

 \begin{eqnarray}
e^{-\varphi_{5}} = \left[\cosh^2(\gamma) -\exp[2(\hu+\hw)] \sinh^2(\gamma){F(x)\over F(y)} \right]^{\alpha_{5}\over 1 + \alpha^2_{5}} , \\
\Phi_{5} = \frac{\tanh(\gamma)}{ \sqrt{1 + \alpha^2_{5}} }\frac{F(y) -F(x)\exp[2(\hu+\hw)] }{ F(y) - F(x)\exp[2(\hu+\hw)]\tanh^2(\gamma) } .
\end{eqnarray}
For $\gamma=0$ this solution corresponds to the distorted five-dimensional black ring. For the distortions field $\hu=\hw=\hv=0$ this solution corresponds to asymptotically flat five dimensional black rings in EMd gravity. The transformation does not affect the distortion
fields $\hu$, $\hv$ and $\hw$. Thus, the transformation electrically charges the black hole
only. In other words, we have a solution which represents a five-dimensional EMdL black
ring distorted by external electrically neutral sources, such that when the distortion fields $\hu$,
$\hw$, and $\hv$ vanish, the solution represents a five-dimensional EMdL black ring solution in an empty, asymptotically flat universe. The metric (\ref{metricEMDL}) is not asymptotically flat due to the presence of sources that distort the black hole. 

The first condition that we wish to formulate is the no-conical regularity condition. The analysis of the no-coniocal singularity of this solution gives us the same result as the case of the uncharged distorted black ring \cite{Abdolrahimi2020}. For the special case where, $\hw=-\hu/2$, we can remove the conical singularity by considering for the distortion fields to satisfy the following condition ($n=0..4$)
\ba\n{Noconical}
\frac{e^{\hv+2\hu+\hw}\rvert_{\theta=\pi}}{e^{\hv+2\hu+\hw}\rvert_{\theta=0}}=\exp\biggl[-\frac{3}{2}(1-\mu)\sum_{n=1}^4 \sum_{k=n}^4 \frac{a_k}{\mu^n}\biggl]=\frac{\sqrt{1+\mu}}{\sqrt{1-\mu}}.
\ea For the dipole distortion ($a_{n>1}=0$) in Eq. (\ref{Noconical}), we get the following relation between $a_1$ and the parameter $\mu$ ,
\be\n{Noconicald}
a_1=\frac{\mu}{3(1-\mu)} \ln\left(\frac{1-\mu}{1+\mu}\right).
\ee For the quadrupole distortion ($a_1=0,a_2\neq 0$, $a_{n>2}=0$) in Eq. (\ref{Noconical}), we get the following relation relation between $a_2$ and the parameter $\mu$:
\be\n{Noconicalq}
a_2=\frac{\mu^2}{3(1-\mu^2)} \ln\left(\frac{1-\mu}{1+\mu}\right) \,. 
\ee
This implies that the conical singularity condition in the distorted charged black ring is removable for special values of multiple moments. 

Also consider the $(y,\psi)$ section of the metric (\ref{MetricA}). We can identify $\psi$ with period $\Delta_\psi=2\pi$, without creating any conical singularity
in our local black hole solution, which is defined in the area near the horizon (recall that infinity is at $x=y=-1$).

\section{Analysis and Summary}
We consider the space-time near the regular horizon of the black hole far away from the sources. In this case, the solution represents a local black hole. We focus on the study of the deformations of the horizon. The distorted black hole horizon is defined by $y\rightarrow -\infty$. The metric of a constant $t$ slice through the horizon (i.e., horizon surface) is
\be
ds^2=\frac{(\cosh\gamma)^{\frac{2}{1+\alpha_5^2}}}{A^2}\biggl(F(x)e^{-2\hw_0}d\psi^2+\mu^2e^{2(\hv_0+\hu_0+\hw_0)}\frac{dx^2}{1-x^2}+\mu^2e^{-2\hu_0}\frac{1-x^2}{F(x)}d\phi^2\biggl). \n{MetricHoriozn1}
\ee
Here, $\hu_0$, $\hw_0$, and $\hv_0$ are calculated on the horizon. Boundary values of the distortion functions $\hu,~\hw$ on the horizon are given by
\ba
&&\hu_0=\hu(x)|_H=\sum_{n\geq0}a_{n}x^n,\\ \n{huHorizon}
&&\hw_0=\hw(x)|_H=\sum_{n\geq0}b_{n}x^n. 
\ea
The boundary value of the distortion function $\hv$ on the horizon is
\ba
\hv_0=\hv(x)|_H=-3\sum_{n\geq 1}^4(a_{n}+b_{n})x^{n}+\sum_{n\geq 1}^2(2a_{2n}+b_{2n})-\sum_{n\geq 1}^4\frac{(a_{n}-b_{n})}{2\mu^n}+C\, . \n{hvHorizon}
\ea
The metric Eq. (\ref{MetricHoriozn1}) is conformal to the case of a distorted black ring presented. One can see that the distortion fields $\hu, \hw$, and $\hv$
are smooth on the black ring horizon.
In what follows, we shall set $C=0$ without loss of generality; solutions with different values of constant $C$ can
be related to this solution by rescaling the periods of $\psi$, $\phi$ and the parameter $A$.  The area of the horizon is 
\be \n{areaeq}
\mathcal{A}_H=\frac{4\pi^2\mu^2}{A^3}\sqrt{1+\mu}\cosh\gamma^{\frac{3}{1+\alpha_5^2}}e^{(V_{0}+2U_{0}+W_{0})|_{x=-1}}\int^{1}_{-1} e^{V_0}dx, 
\ee
Note that in the previous paper this expression had a mistake \cite{Abdolrahimi2020}. For $\gamma=0$ this reduces to the correct expression for the area of the horizon. For the case $\hw=-\hu/2$  and for $a_{n\ge2}=0$ we have:
\be
\mathcal{A}_H=\frac{8\pi^2\mu^2}{A^3}\sqrt{1+\mu}\cosh\gamma^{\frac{3}{1+\alpha_5^2}}\frac{e^{3a_1}-1}{3a_1}e^{\frac{3(a_0-a_1(1+\mu))}{2\mu}} \, .
\ee For the case that $a_1\ll 1$, we get

\be
\mathcal{A}_H=\frac{8\pi^2\mu^2}{A^3}\sqrt{1+\mu}\cosh\gamma^{\frac{3}{1+\alpha_5^2}} e^{\frac{3a_0}{2}} \{1- \frac{3a_1}{2\mu} \} \, .
\ee For the case that $a_1=0$, $a_{n\ge3}=0$, and $a_2\ll 1$, we have

\be
\mathcal{A}_H=\frac{8\pi^2\mu^2}{A^3}\sqrt{1+\mu}\cosh\gamma^{\frac{3}{1+\alpha_5^2}} e^{\frac{3a_0}{2}} \{1+ \frac{(5\mu^2-3)a_2}{2\mu^2} \} \, .
\ee Here we should clarify that for the undistorted case (i.e. $a_{n}=0=b_{n}$) the Eq. (\ref{areaeq}) does not match the Eq. (\ref{areaeq1}) because we consider different period for the $\psi$ coordinate.

The surface gravity of a distorted charged black ring is given by
\be\label{sg1}
\kappa=\frac{A}{2\mu}\left(\cosh\gamma\right)^{\frac{-3}{\alpha_5^2+1}}\exp\left[\sum_{n=1}^2 2^{n-1}\left(\frac{A^2}{\alpha\mu}\right)^n[b_n(c_1^n+c_2^n+c_3^n)-a_n(c_1^n-2c_2^n-2c_3^n)]\right]~. 
\ee
Using the no-conical singularity condition we get,
\be\label{sg2}
\kappa=\frac{A}{2\mu}\left(\cosh\gamma\right)^{\frac{-3}{\alpha_5^2+1}}\exp\left[\sum_{n=1}^2 \frac{-3a_n(\mu^{-n}-(-1)^n-1)}{4}\right]~.  
\ee Next, let us consider only the dipole-quadrupole distortions, we get the following for the rescaled surface gravity,
\be\label{sg3}
\tilde{\kappa}=\frac{\kappa}{\kappa_{iso}}=\exp\left[\frac{3(2\mu^2a_2-a_1\mu-a_2)}{4\mu^2}\right] \, ,
\ee where, $\kappa_{iso}=\frac{A}{2\mu}\left(\cosh\gamma\right)^{\frac{-3}{\alpha_5^2+1}}$ is the surface gravity of the undistorted charged black ring. 
\begin{center} 
\begin{figure}[t]
\centering
  \includegraphics[width=7cm]{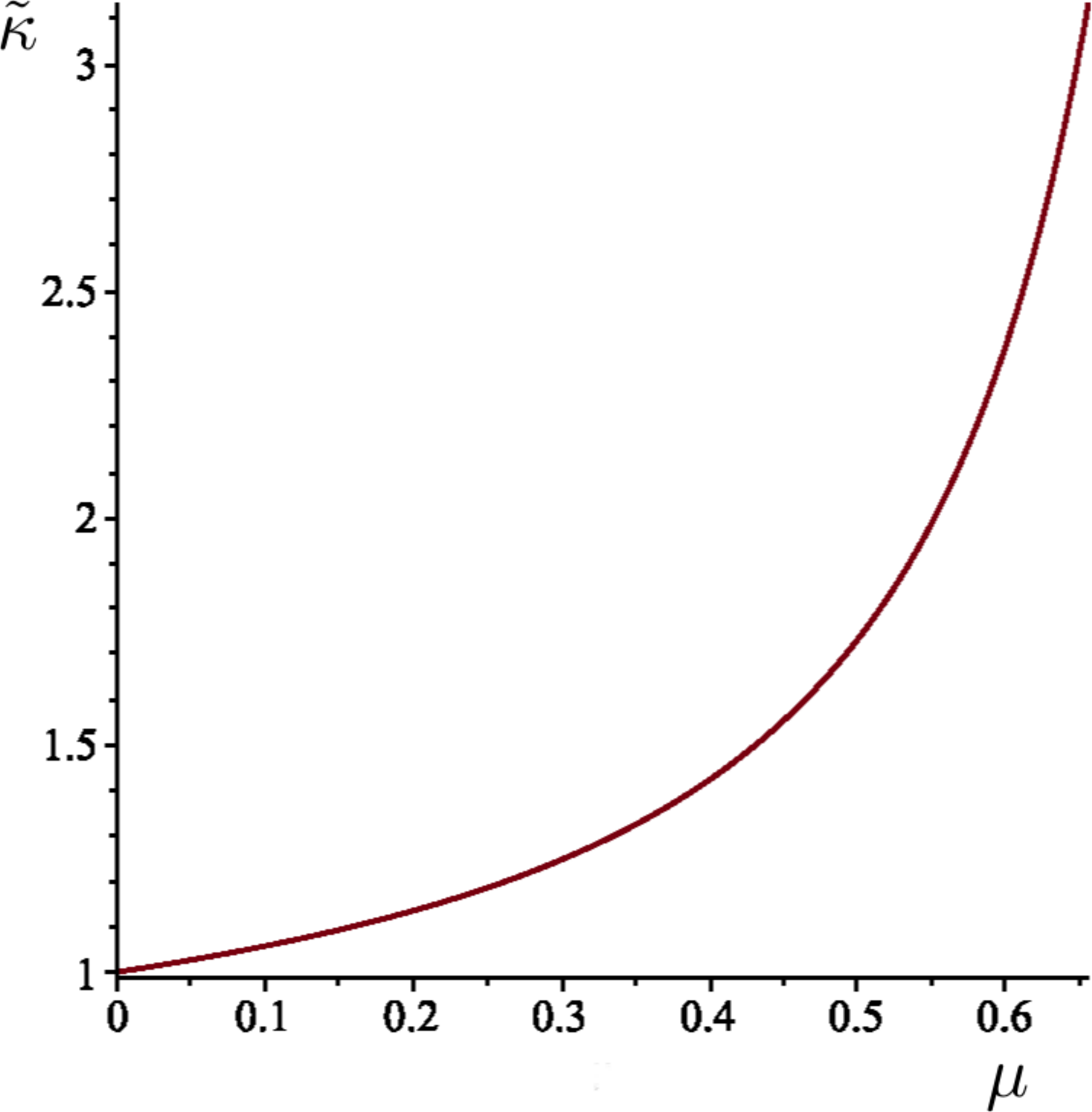}~~~~~~~
   \includegraphics[width=7cm]{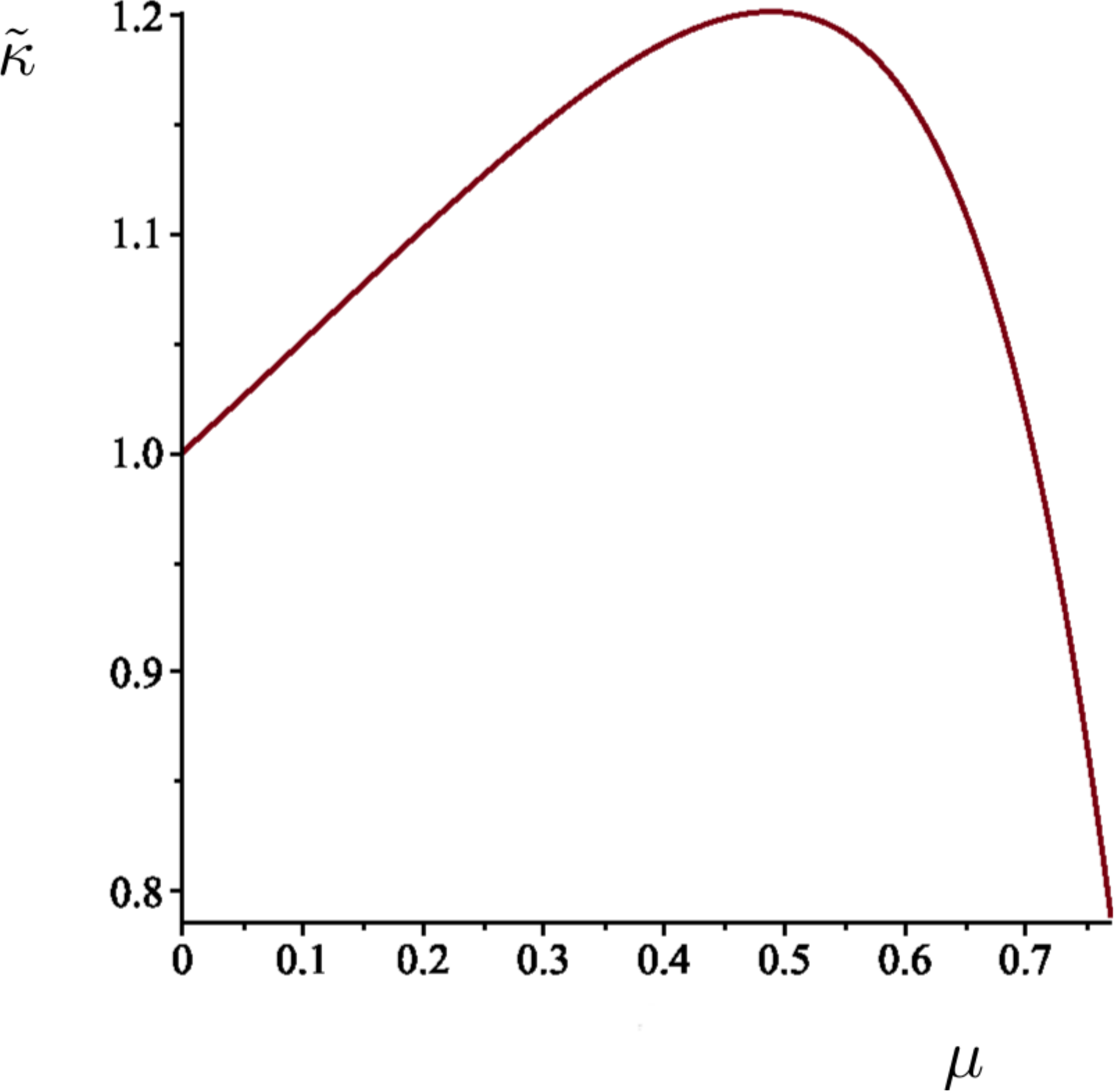}
   \caption{Behavior of $\tilde{\kappa}$ as a function of parameter $\mu$. On the left we have plotted the behavior of $\tilde{\kappa}$ for dipole multiple moment with $\mu \in (0..0.656)$. On the right we have plotted the behavior of $\tilde{\kappa}$ for quadrupole multiple moment with $\mu \in (0..0.771)$. We choose multiple moments (dipole or quadrupole) to be less than $-1$. \label{kappa}}
\end{figure}
\end{center} 
\begin{center}
\begin{figure}[t]
\centering
  \includegraphics[width=7cm]{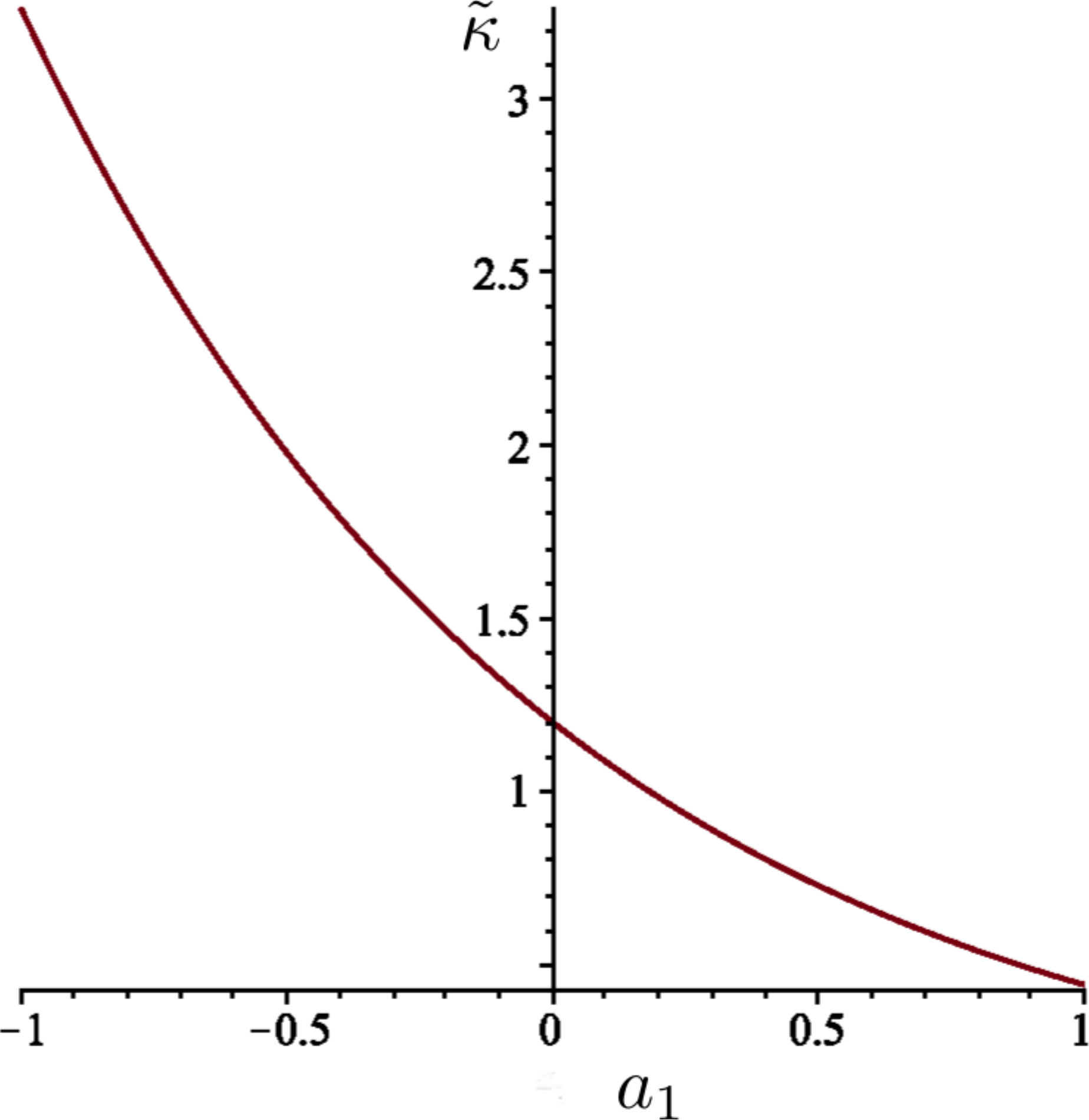}~~~~~~~
   \includegraphics[width=7cm]{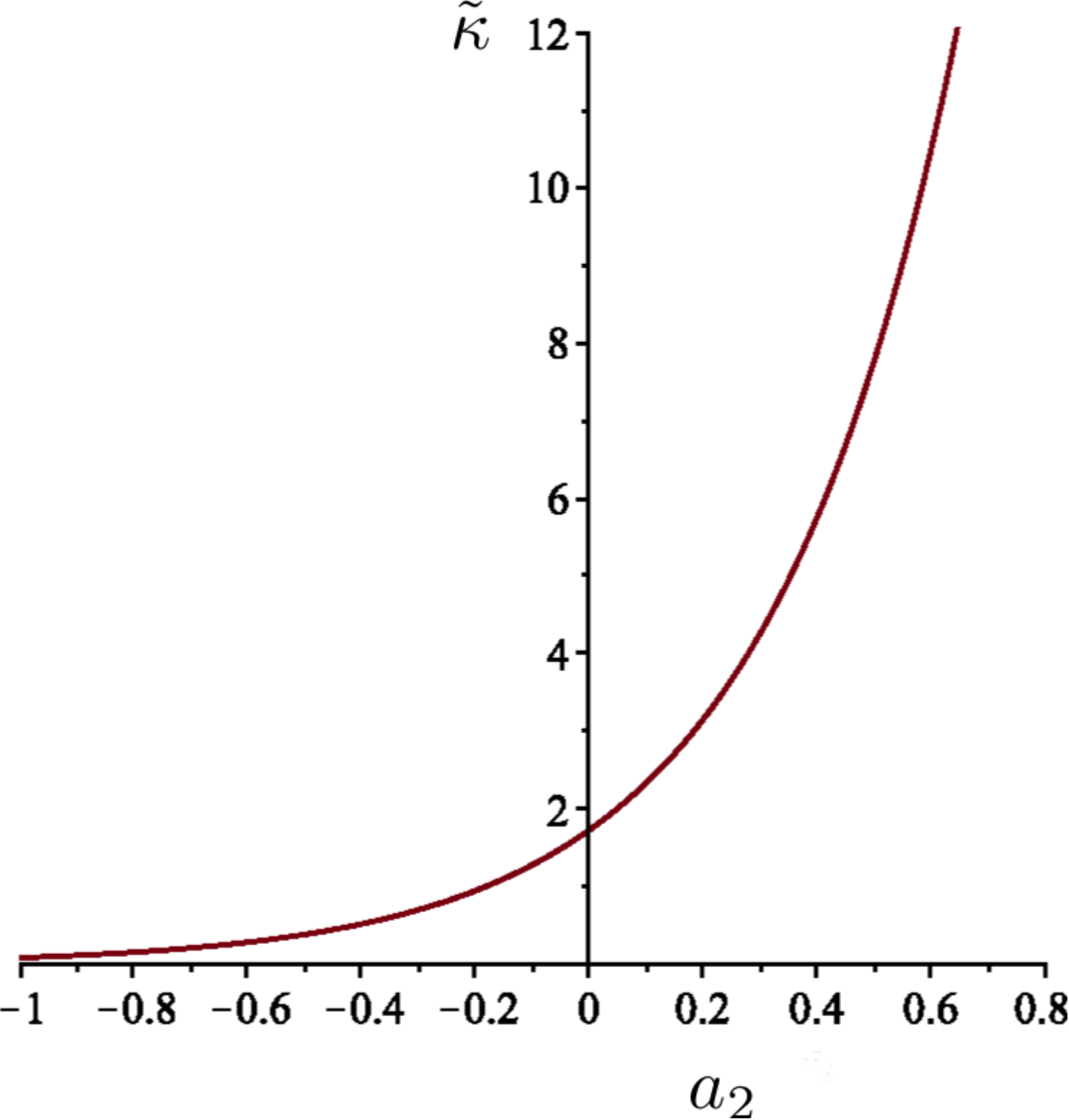}
   \caption{On the left we have plotted the behavior of $\tilde{\kappa}$ as a function of the dipole multiple moment $a_1$ for $\mu=0.5$. On the right we have plotted the behavior of $\tilde{\kappa}$ as a function of the quadrupole multiple moment $a_2$ for $\mu=0.5$.\label{kappa2}}
\end{figure}
\end{center}
In the Fig. \ref{kappa}, we have plotted the behavior of $\tilde{\kappa}=\kappa/\kappa_{iso}$ as a function of the parameter $\mu$, where  $\mu$ is a parameter of the background undistorted black ring. Relations (\ref{Noconicald}) and (\ref{Noconicalq}) yield a prescription for  the multiple moments in terms of $\mu$, fine-tuned to avoid conical singularity. We consider $|a_n|<1$ and $|b_n|<1$. The reason for this is that physically, a small multiple moment corresponds to distortion of the black object distorted by masses smaller than the object itself. On the other hand, a large multiple moment would result from surrounding the black object with masses much larger than the object itself. In our analysis, we restrict ourselves, to small multiple moment values, so that the resulting fine tuned multiple moments are likewise small, i.e., less than $-1$ (large values of $\mu$ imply correspondingly large multiple moment values using (\ref{Noconicald}) and (\ref{Noconicalq}) ). We consider $\mu \in (0,0.656)$ for the dipole multiple moment. As for the quadrupole multiple moment, we consider $\mu \in (0,0.771)$. In the left diagram of Fig. \ref{kappa}, we consider the case of dipole multiple moment by setting $a_2=0$ and use the Eq. (\ref{Noconicald}) to related $a_1$ with the parameter $\mu$. Note that increasing $\mu$ implies increasingly negative $a_1$.

In the right diagram of Fig. \ref{kappa}, we consider a quadrupole multiple moment by setting $a_1=0$, and use the Eq. (\ref{Noconicalq}) to related $a_2$ with the parameter $\mu$. Moreover, if one increases $\mu$ then $a_2$ is going to be increased negatively. We see that an increasingly large dipole moment strengthens the surface gravity of the distorted black ring. An increasingly large quadrupole moment strengthens the surface gravity of the distorted black ring. This happens only up to a certain maximal value. Beyond this value the surface gravity of the black ring decreases, ultimately becoming less than its undistorted counterpart as the quadrupole parameter attains its maximal negative value. For $\mu\rightarrow 1$, the surface gravity $\kappa\rightarrow 0$. However, note that this case corresponds to $|a_2|\rightarrow\infty$. We don't analyze this case since we consider $|a_n|<1$. 

Next, let us consider a distorted charged black ring having both dipole and quadrupole distortions. In  Fig. \ref{kappa2}, we have plotted the behavior of $\tilde{\kappa}=\kappa/\kappa_{iso}$ as a function of the multiple moments for a fixed parameter $\mu=0.5$ in the case of dipole-quadrupole.  For the case of  dipole-quadrupole distortion, we have two cases. In the left diagram of Fig. \ref{kappa2}, we fine tune  $a_2$, leaving $a_1$ as a free parameter. We plot the behavior of $\tilde{\kappa}$ as a function of the dipole multiple moment $a_1$ by replacing
\be\n{a2}
a_2 =-\frac{ \ln\left(\frac{1+\mu}{1-\mu}\right)\mu^4-3\mu^4a_1+3\mu^3a_1}{3\mu^2(1-\mu^2)}\, ,
\ee which is derived from the no conical singularity condition (\ref{Noconical}) for the joint dipole-quadrupole case. In the right diagram of Fig. \ref{kappa2},  we have fine tuned $a_1$ and plotted the behavior of $\tilde{\kappa}$ as a function of the quadrupole multiple moment $a_2$ by inverting (\ref{a2})
\be
a_1 =-\frac{ \ln\left(\frac{1+\mu}{1-\mu}\right)\mu^2-3\mu^2a_2+3a_2}{3\mu(1-\mu)}\, .
\ee We find, regardless of the value of $\mu$, that behavior of $\tilde{\kappa}$ remains the same. Namely, as it is respectively illustrated in the left and right diagrams of Fig. \ref{kappa2}, $\tilde{\kappa}$ monotonically decreases for positive values of dipole moment $a_1$ and monotonically increases for positive values of quadrupole multiple moment $a_2$.

In summary, we have constructed the metric of a distorted five-dimensional charged static black ring, representing a local charged black ring distorted by external static and neutral distributions of matter sources. The solution is free of singularities, with distortion fields $\hu$, $\hw$ and $\hv$ that are regular on the horizon. We observe that by careful fine-tuning of the external distorting sources we can have a solution that is free of the conical singularity that is present in the isolated charged black ring. Similar effect was illustrated for uncharged distorted black ring \cite{Abdolrahimi2020}. We have also analyzed the behavior of the distorted black ring’s surface gravity. For quadrupole distortions increasingly large (negative) dipole distortions weaken the surface gravity of the black ring, and increasingly large (positive) quadrupole distortions strengthen the surface gravity of the black ring. The surface gravity of the distorted black hole can be significantly larger than the undistorted charged black ring one. For further analysis, it would be interesting to study the effects of distortions on the shape of the horizon.

\section*{Acknowledgments}
S.A. appreciated the hospitality of the KITP, as a KITP scholar. This research was supported in part by the National Science Foundation under Grant No. NSF PHY-1748958.


\begin{thebibliography}{}
\bibitem{Abdolrahimi2020} S. Abdolrahimi, R. B. Mann, and C. Tzounis, ``Distorted black ring",  Phys. Rev. D {\bf101}, 10, 104002 (2020).  
\bibitem{GerochH} R.~Geroch and J.~B.~Hartle, ``Distorted Black Holes," J. Math. Phys. {\bf 23}, 680 (1982).
\bibitem{Weyl} H.~Weyl, ``On the Theory of Gravitation," Ann. Phys., {\bf 54}, 117 (1917).
\bibitem{dis1} W.~Israel and K.~A.~Khan, ``Collinear Particles and Bondi Dipoles in General Relativity," Nuovo Cimento {\bf 33}, 331 (1964).
\bibitem{dis2} L. A. Mysak and G. Szekeres, ``Behavior of the Schwarzschild singularity in superimposed gravitational fields", Can. J. Phys. {\bf 44} (1966) 617. 
 \bibitem{dis3} W.~Israel, ``Event Horizons in Static Vacuum Space-Times," Phys. Rev. {\bf164}, 1776 (1967).
 \bibitem{dis4} S.~Chandrasekhar, {\it The Mathematical Theory of Black Holes}, Clarendon Press, Oxford University Press, New York, Oxford, (1992), p. 583.
\bibitem{dis5} 
  V.~P.~Frolov and A.~A.~Shoom,
  ``Interior of Distorted Black Holes,''
  Phys.\ Rev.\ D {\bf 76}, 064037 (2007). 
 \bibitem{Fairhurst} S.~Fairhurst and B.~Krishnan, ``Distorted Black Holes with Charge," Int. J. Mod. Phys. D {\bf 10}, 691 (2001).
  \bibitem{Israel1776} W. Israel, ``Event horizons in static vacuum space-times", Phys. Rev. {\bf 164} (1967) 1776. 
  \bibitem{Abdolrahimi2009}S. Abdolrahimi, V. P. Frolov and A. A. Shoom, ``Interior of a Charged Distorted Black Hole,'' Phys. Rev. D, {\bf 80}, 024011(2009).
 \bibitem{Tomimatsu} A.~Tomimatsu, ``Distorted Rotating Black Holes," Phys. Lett. A {\bf 103}, 374 (1984).
\bibitem {Breton1997} N.~Bret\'{o}n, T.~Denisova, and V.~Manko, ``A Kerr Black Hole in the External
Gravitational Field," Phys.\ Lett.\ A {\bf230}, 7 (1997).
\bibitem {Breton1998}N.~Bret\'{o}n, A. A. Garc\'ia, V. S. Manko, and T. E. Denisova, Arbitrarily deformed Kerr
Newman black hole in an external gravitational field, Phys. Rev. D 57 (1998) 3382.
\bibitem{Ansorg2008} M. Ansorg and J. Hennig, ``The Inner Cauchy horizon of axisymmetric and stationary black
holes with surrounding matter", Class. Quant. Grav. {\bf 25} (2008) 222001 [arXiv:0810.3998].
\bibitem{Ansorg2009} M. Ansorg and J. Hennig, ``Inner Cauchy Horizon of Axisymmetric and Stationary Black Holes with Surrounding Matter in Einstein-Maxwell Theory", Phys. Rev. Lett. {\bf102} (2009) 221102 [arXiv:0903.5405].
\bibitem{Ansorg2009-2} J. Hennig and M. Ansorg, ``The inner Cauchy horizon of axisymmetric and stationary black holes with surrounding matter in Einstein-Maxwell theory: study in terms of soliton methods", Annales Henri Poincare {\bf10} (2009) 1075 [arXiv:0904.2071].
 \bibitem{GeneralizedWeyl}  R.~Emparan and H.~S.~Reall,
  ``Generalized Weyl Solutions,''
  Phys.\ Rev.\ D {\bf 65}, 084025 (2002).
   \bibitem{Abdolrahimi2010} S. Abdolrahimi, A. A. Shoom, and D. N. Page, Distorted 5-dimensional vacuum black hole , Phys. Rev. D {\bf 82} (2010) 124039 [arXiv:1009.5971].
 \bibitem{Abdolrahimi2013} 
  S.~Abdolrahimi and A.~A.~Shoom,
  ``Distorted Five-dimensional Electrically Charged Black Holes,''
  Phys.\ Rev.\ D {\bf 89}, no. 2, 024040 (2014).
    \bibitem{Abdolrahimi2014} 
  S.~Abdolrahimi, J.~Kunz and P.~Nedkova, ``Myers-Perry Black Hole in an External Gravitational Field,''
  Phys.\ Rev.\ D {\bf 91}, 064068 (2015).
\bibitem{SCN3} A.~G.~Doroshkevich, Ya.~B.~Zel'dovich, and I.~D.~Novikov, ``Gravitational Collapse of Nonsymmetric and Rotating Masses," JETP {\bf 22}, 122 (1996).
\bibitem{dis2A} P.~Peters, ``Toroidal Black Boles?" J. Math. Phys. {\bf 20}, 1481 (1979). 
\bibitem{dis4A} B.~Xanthopoulos, ``Local Toroidal Black Holes That are Static and Axisymmetric," Proc. R. Soc. Lond. A {\bf 388}, 17 (1983).
\bibitem{Poisson:2009qj} 
  E.~Poisson and I.~Vlasov,
  ``Geometry and dynamics of a tidally deformed black hole,''
  Phys.\ Rev.\ D {\bf 81}, 024029 (2010).
  \bibitem{Shoom:2015rda} 
  A.~A.~Shoom,
  ``Distorted stationary rotating black holes,''
  Phys.\ Rev.\ D {\bf 91}, no. 6, 064030 (2015). 
    \bibitem{Semerak:2016gfz} 
  O.~Semerák and M.~Basovník,
  ``Geometry of deformed black holes. I. Majumdar-Papapetrou binary,''
  Phys.\ Rev.\ D {\bf 94}, no. 4, 044006 (2016).
  \bibitem{Basovnik:2016awa} 
  M.~Basovn\`{i}k and O.~Semer\'{a}k,
  ``Geometry of deformed black holes. II. Schwarzschild hole surrounded by a Bach-Weyl ring,''
  Phys.\ Rev.\ D {\bf 94}, no. 4, 044007 (2016).
   \bibitem{Ansorg:2010ru} 
  M.~Ansorg, J.~Hennig and C.~Cederbaum,
  ``Universal properties of distorted Kerr-Newman black holes,''
  Gen.\ Rel.\ Grav.\  {\bf 43}, 1205 (2011).
   \bibitem{Shoom:2015slu} 
  A.~A.~Shoom, C.~Walsh and I.~Booth,
  ``Geodesic motion around a distorted static black hole,''
  Phys.\ Rev.\ D {\bf 93}, no. 6, 064019 (2016).
 \bibitem{Abdolrahimi:2014msa} 
  S.~Abdolrahimi,
  ``Thermodynamic of Distorted Reissner-Nordström Black Holes in Five-Dimensions,''
  Springer Proc.\ Phys.\  {\bf 170}, 217 (2016).
   \bibitem{Abdolrahimi:2017pmt} 
  S.~Abdolrahimi, J.~Kunz and P.~Nedkova,
  ``Rotating distorted black holes in higher dimensions,'' Proceedings of the MG14 Meeting on General Relativity, University of Rome “La Sapienza”, Italy, 12 – 18 July 2015, pp. 1763-1768 (2017).
  \bibitem{Kunz:2017mfe} 
  J.~Kunz, P.~Nedkova and S.~Yazadjiev,
  ``Magnetized Black Holes in an External Gravitational Field,''
  Phys.\ Rev.\ D {\bf 96}, no. 2, 024017 (2017).
 \bibitem{Pilkington} T. Pilkington, A. Melanson, J. Fitzgerald, and I. Booth, Trapped and ``marginally trapped surfaces in Weyl-distorted Schwarzschild solutions", Class. Quant. Grav. {\bf 28} (2011) 125018 [gr-qc/1102.0999].
  \bibitem{Abdolrahimi:2015c} 
  S.~Abdolrahimi, J.~Kunz, P.~Nedkova and C.~Tzounis,
  ``Properties of the distorted Kerr black hole,''
  JCAP {\bf 1512}, 009 (2015). 
   \bibitem{Abdolrahimi:2015a} 
  S.~Abdolrahimi, R.~B.~Mann and C.~Tzounis,
  ``Distorted Local Shadows,''
  Phys.\ Rev.\ D {\bf 91}, no. 8, 084052 (2015). 
   \bibitem{Abdolrahimi:2015b} 
  S.~Abdolrahimi, R.~B.~Mann and C.~Tzounis,
  ``Double Images from a Single Black Hole,''
  Phys.\ Rev.\ D {\bf 92}, 124011 (2015). 
   \bibitem{Grover:2018tbq} 
  J.~Grover, J.~Kunz, P.~Nedkova, A.~Wittig and S.~Yazadjiev,
  ``Multiple shadows from distorted static black holes,''
  Phys.\ Rev.\ D {\bf 97}, no. 8, 084024 (2018). 
  \bibitem{ER3}
  R.~Emparan and H.~S.~Reall,
  ``Black Holes in Higher Dimensions,''
  Living Rev.\ Rel.\  {\bf 11} (2008) 6.
   \bibitem{Emparan2008}  R.~Emparan and H.~S.~Reall,
  ``Black Holes in Higher Dimensions,''
  Living Rev.\ Rel.\  {\bf 11}, 6 (2008).
  \bibitem{ER2} R.~Emparan and H.~S.~Reall,
 ``A Rotating black ring solution in five-dimensions,''
  Phys.\ Rev.\ Lett.\  {\bf 88}, 101101 (2002). 
  \bibitem{Elvang:2007rd} 
  H.~Elvang and P.~Figueras,``Black Saturn,''
  JHEP {\bf 0705}, 050 (2007).
  \bibitem{Iguchi} H. Iguchi and T. Mishima, ``Black diring and infinite nonuniqueness," Phys. Rev. D 75, 064018 (2007). 
\bibitem{Elvang2008} H. Elvang, M. Rodriguez, ``Bicycling black rings," J. High Energy Phys. 4:045 (2008). 
\bibitem{Izumi2008} K. Izumi, ``Orthogonal black di-ring solution," Prog. Theor. Phys. 119:757-774 (2008). 
\bibitem{Evslin:2007fv} 
  J.~Evslin and C.~Krishnan,
  ``The Black Di-Ring: An Inverse Scattering Construction,''
  Class.\ Quant.\ Grav.\  {\bf 26}, 125018 (2009). 
  \bibitem{Emparan:2011br}
  R.~Emparan,
  ``Blackfolds,''
  arXiv:1106.2021. 
\bibitem{Papadopoulos1984} D. Papadopoulos, B. Xanthopoulos, ``Local black holes are type D on the horizon", Il Nuovo Cimento B 83 (1984) 113.
  \bibitem{Ashtekar1999} A. Ashtekar, C. Beetle and S. Fairhurst, ``Isolated horizons: a generalization of black hole mechanics", Class. Quantum Grav. 16 L1 (1999).
 \bibitem{Ashtekar2000} A. Ashtekar, C. Beetle and S. Fairhurst, ``Mechanics of isolated horizons", Class. Quantum Grav. 17 253-298 (2000).
\bibitem{Ashtekar2000v2} A. Ashtekar, S. Fairhurst and B. Krishnan, ``Isolated horizons: Hamiltonian evolution and the first law?, Phys. Rev. D 62, 104025 (2000).
  \bibitem{MTW} C.~W.~Misner, K.~S.~Thorne and J.~A.~Wheeler, {\em Gravitation},   W. H. Freeman and Co., San Francisco, (1973).
  \bibitem{Sen}A. Sen, ``Electric magnetic duality in string theory", Nucl. Phys. B {\bf 404}, 1993. 
\bibitem{Gibbons} G. W. Gibbons and K.-i. Maeda, “Black Holes and
Membranes in Higher Dimensional Theories with Dila-
ton Fields,” Nucl. Phys. B{\bf298}, 741–775 (1988).
\bibitem{Garfinkle} D. Garfinkle, G. T. Horowitz, and A. Strominger, “Charged black holes in string theory,” Phys. Rev. D {\bf 43},
3140 (1991), [Erratum: Phys. Rev.D45,3888(1992)].
 \bibitem{Yazadjiev} Stoytcho S. Yazadjiev, ``Asymptotically and non-asymptotically flat static
black rings in charged dilaton gravity", arXiv:hep-th/0507097 (2005). 
\bibitem{Multipoles1}J. D. Jackson, ``Classical Electrodynamics" (John Wiley \& Sons, Inc., New York, 1975), p. 90-93.
\bibitem{Multipoles2} K. I. Ramachandran, G. Deepa, and K. Namboori, Computational Chemistry and Molecular Modeling: Principles and Applications (Springer-Verlag Berlin Heidelberg, 2008), p. 258.
\bibitem{KL} H.~Kunduri~and~J.~Lucietti, ``Electrically charged dilatonic black rings", Phys. Lett. {\bf B609},143 (2005).

  
  
 \end{thebibliography}
\end{document}